\documentclass[runningheads]{svmult}

\usepackage{makeidx}   % allows index generation
\usepackage{graphicx}  % standard LaTeX graphics tool
\usepackage{subeqnar}  % subnumbers individual equations
\usepackage{multicol}  % used for the two-column index
\usepackage{physprbb}  % modified textarea for proceedings,
\makeindex             % used for the subject index
\begin{document}
\title*{Weak energy condition violation and
superluminal travel}
\toctitle{Weak energy condition violation and superluminal travel}
% allows explicit linebreak for the table of content
%
%

\author{Francisco Lobo\thanks{Email address: flobo@cosmo.fis.fc.ul.pt}
\inst{} and Paulo Crawford\thanks{Email address:
crawford@cosmo.fis.fc.ul.pt}\inst{}}
\authorrunning{Francisco Lobo and Paulo Crawford}
% if there are more than two authors,
% please abbreviate author list for running head
%
%
\institute{\em Centro de Astronomia e Astrof\'{\i}sica
 da Universidade de Lisboa \\
 Campo Grande, Ed. C8 1749-016 Lisboa, Portugal}

\maketitle

\begin{abstract}

Recent solutions to the Einstein Field Equations involving
negative energy densities, i.e., matter violating the
weak-energy-condition, have been obtained, namely traversable
wormholes, the Alcubierre warp drive and the Krasnikov tube. These
solutions are related to superluminal travel, although locally the
speed of light is not surpassed. It is difficult to define
faster-than-light travel in generic space-times, and one can
construct metrics which apparently allow superluminal travel, but
are in fact flat Minkowski space-times. Therefore, to avoid these
difficulties it is important to provide an appropriate definition
of superluminal travel.

We investigate these problems and the relationship between
weak-energy-condition violation and superluminal travel.

\end{abstract}

\section{Introduction}
Much interest has been revived in superluminal travel in the last
few years. Despite the term {\it superluminal}, it is not possible
to travel faster than the speed of light, {\it locally}. The point
to note is that one can make a round trip, between two points
separated by a distance $D$, in an arbitrarily short time as
measured by an observer that remained at rest at the starting
point, by varying one's speed or by changing the distance one is
to cover.

Apart from wormholes {\cite{Morris,Visser}}, two spacetimes which
allow superluminal travel are the Alcubierre warp drive
\cite{Alcubierre} and the solution known as the Krasnikov tube
\cite{Krasnikov,Everett}. These spacetimes suffer from a severe
drawback, as they require negative energy densities or {\it
exotic} matter, i.e., they violate the weak energy condition
(WEC). In fact, they violate all the known energy conditions and
averaged energy conditions, which are fundamental to the
singularity theorems and theorems of classical black hole
thermodynamics \cite{Visser}. Although classical forms of matter
obey these energy conditions, it is a well-known fact that they
are violated by certain quantum fields.

One is liable to ask if it is possible to have superluminal travel
without the violation of the WEC. But it's fundamental, first, to
provide an adequate definition of {\it superluminal travel}, which
is no trivial matter \cite{VB,VBL}. A plausible and general idea
is that the modification of the metric would allow the propagation
of signals between two spacetime points, that otherwise would be
causally disconnected.

The aim of this work is to investigate whether it is possible to
have superluminal travel, without the violation of the WEC. For
self-consistency and self-completeness, we present in this article
an overview of the basics of the above-mentioned solutions and the
analysis of an important theorem produced by Ken Olum \cite{Olum}.
We also briefly outline a new form of constraint, designated by
the Quantum Inequality, deduced from quantum field theory by Ford
and Roman \cite{F&R1}. The present work serves as a bridge to
ongoing research on spacetimes which generate closed timelike
curves.

\section{Warp drive basics}

Within the framework of general relativity, it is possible to warp
spacetime in a small {\it bubblelike} region, in such a way that
the bubble may attain arbitrarily large velocities. Inspired in
the inflationary phase of the early Universe, the enormous speed
of separation arises from the expansion of spacetime itself. The
model for hyperfast travel is to create a local distortion of
spacetime, producing an expansion behind the bubble, and an
opposite contraction ahead of it.

Consider a bubble moving along the $Oz$ axis with velocity, $v$.
Therefore, the Alcubierre spacetime metric, in cylindrical
coordinates, is given by (with the notation $G=c=1$):
\begin{equation}
ds^2=-dt^2+d\rho^2+\rho^2 d\phi^2+(dz-vfdt)^2
\end{equation}
where:
\begin{eqnarray*}
v(t)&=&\frac{dz_0(t)}{dt}\\
r(t)&=&[\rho^2+(z-z_{0})^2]^{1/2}
\end{eqnarray*}
and the form function, $f(r)$, is given by \cite{Alcubierre}:
\[
f(r)=\frac{\tanh(\sigma(r+R))-\tanh(\sigma(r-R))}{2\tanh(\sigma
R)}
\]
in which $R>0$ and $\sigma>0$ are two arbitrary parameters. $R$ is
the radius of the bubble, and $\sigma$ can be interpreted as being
inversely proportional to the bubble wall thickness.

Notice that for large $\sigma$, the form function rapidly
approaches a {\it top hat} function:
\[
\lim_{\sigma \rightarrow \infty} f(r)=\left\{ \begin{array}{ll}
1, & {\rm if}\; r\in[-R,R],\\
0, & {\rm if}\; otherwise.
\end{array}
\right.
\]

\subsection{The expansion of the volume elements}

The expansion of the volume elements is given by:
\begin{equation}
\theta=v\frac{z-z_0}{r}\frac{df(r)}{dr}
\end{equation}

Consider a spaceship immersed within the bubble. The center of the
perturbation corresponds to the spaceship's position, $z_0(t)$.
The volume elements are expanding behind the spaceship, and
contracting in front of it.

\begin{figure}[t]
\centering
\includegraphics[width=9cm]{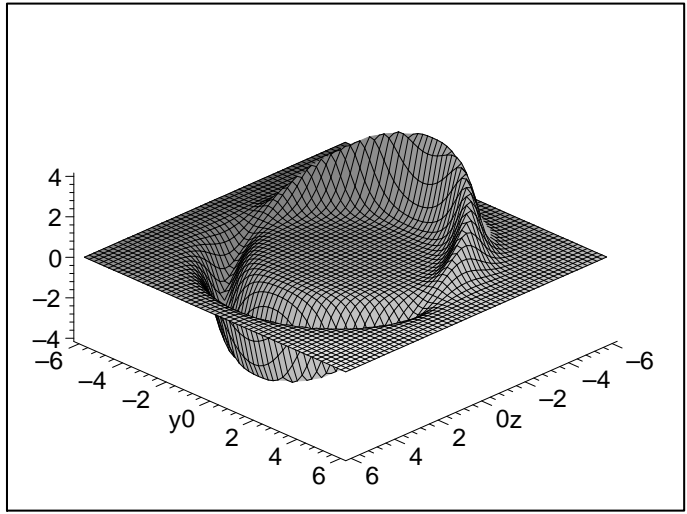}
\caption{The expansion of the volume elements. These are expanding
behind the spaceship, and contracting in front of it.
\label{fig1}}
\end{figure}

Note that the spaceship moves along a timelike curve, regardless
of the value of $v(t)$. To verify this statement we simply
substitute $z=z_0(t)$ in the metric, eq.$(1)$, which reduces to:
\begin{equation}
d\tau=dt
\end{equation}
From which we conclude that the proper time equals the coordinate
time, therefore the spaceship suffers no time dilation effects
during it's motion. It is also not difficult to prove that the
spaceship moves along a geodesic.

\subsection{Superluminal travel in the warp drive}

To demonstrate that it is possible to travel to a distant point
and back in an arbitrary short time interval, let us consider two
distant stars, $A$ and $B$, separated by a distance $D$ in flat
spacetime. Suppose that, at the instant $t_0$, a spaceship
initiates it's movement using the engines, moving away from $A$
with a velocity $v<1$. It comes to rest at a distance $d$ from
$A$. For simplicity, assume that $R\ll d\ll D$.

It is at this instant that the perturbation of spacetime appears,
centered around the spaceship's position. The perturbation pushes
the spaceship away from $A$, rapidly attaining a constant
acceleration, $a$. Half-way between $A$ and $B$, the perturbation
is modified, so that the acceleration rapidly varies from $a$ to
$-a$. The spaceship finally comes to rest at a distance, $d$, from
$B$, in which the perturbation disappears. It then moves to $B$ at
a constant velocity in flat spacetime. The return trip to $A$ is
analogous.

If the variations of the acceleration are extremely rapid, the
total coordinate time, $T$, in a one-way trip will be:
\[
T=2\left( \frac{d}{v}+\sqrt{\frac{D-2d}{a}} \right)
\]
The proper time of the stars are equal to the coordinate time,
because both are immersed in flat spacetime. The proper time
measured by observers within the spaceship is given by:
\[
\tau=2\left( \frac{d}{\gamma v}+\sqrt{\frac{D-2d}{a}} \right)
\]
with $\gamma =(1-v^2)^{-1/2}$. The time dilation only appears in
the absence of the perturbation, in which the spaceship is moving
with a velocity $v$, using only it's engines in flat spacetime.

Using $R\ll d\ll D$, we can then obtain the following
approximation:
\[
\tau\approx T\approx 2\sqrt{\frac{D}{a}}
\]
We verify that $T$ can be made arbitrarily short, increasing the
value of $a$. The spaceship may travel faster than the speed of
light. However, it moves along a spacetime temporal trajectory,
contained within it's light cone, for light suffers the same
distortion of spacetime \cite{Alcubierre}.

\subsection{The violation of the WEC}

Given a stress energy tensor $T_{\mu\nu}$, and a timelike vector
$U^{\mu}$, the WEC states:
\begin{equation}
T_{\mu\nu}U^{\mu}U^{\nu}\geq0
\end{equation}
This condition is equivalent to the assumption that any timelike
observer measures a local positive energy density.

\begin{figure}
\centering
\includegraphics[width=8cm]{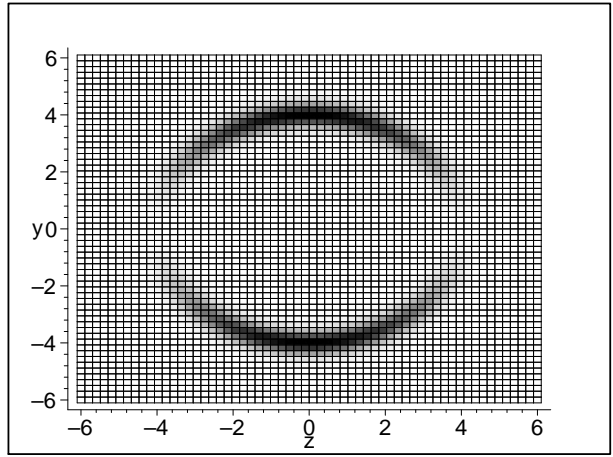}
\caption{The negative energy density for a longitudinal cross
section of the Alcubierre bubble. The energy density is
distributed in a toroidal region perpendicular to the direction of
travel. We have considered the following values: $v=2$, $\sigma
=2$ and $R=8$. \label{fig2}}
\end{figure}

We verify that for the warp drive metric, the WEC is violated:
\begin{equation}
T_{\mu\nu}U^{\mu}U^{\nu}=-\frac{1}{32\pi}\frac{v^2
\rho^2}{r^2}\left( \frac{df}{dr} \right)^2<0
\end{equation}

It is also possible to show that the dominant (DEC) and the strong
energy condition (SEC) are also violated. In fig.$2$ we verify
that the distribution of the negative energy density is
concentrated in a toroidal region perpendicular to the direction
of travel.

\subsection{Interesting aspects of the Alcubierre spacetime}

\subsubsection{The Krasnikov analysis:}

Krasnikov discovered a fascinating aspect of the warp drive, in
which an observer on a spaceship cannot create nor control on
demand an Alcubierre bubble, with $v>c$, around the ship
\cite{Krasnikov}. It is easy to understand this, as an observer at
the origin (with $t=0$), cannot alter events outside of his future
light cone, $|r|\leq t$, with $r=(\rho^2+z^2)^{1/2}$. Applied to
the warp drive, points on the outside front edge of the bubble are
always spacelike separated from the centre of the bubble.

The analysis is simplified in the proper reference frame of an
observer at the centre of the bubble. Using a transformation,
$z'=z-z_{0}(t)$, the metric is given by:
\begin{equation}
ds^2=-dt^2+d\rho^2+\rho^2 d\phi^2+(dz'+(1-f)vdt)^2
\end{equation}

Consider a photon emitted along the $+Oz$ axis (with
$ds^2=d\rho=0$):
\begin{equation}
\frac{dz'}{dt}=1-(1-f)v
\end{equation}

Initially, the photon has $\frac{dz'}{dt}=1$ (because $f=1$ in the
interior of the bubble). However, at some point $z'=z'_c$, with
$f=1-\frac{1}{v}$, we have $\frac{dz'}{dt}=0$ \cite{Everett}. Once
photons reach $z'_c$, they remain at rest relative to the bubble
and are simply carried along with it. This behaviour is
reminiscent of an {\it event horizon}.

\subsubsection{Reminiscence of an Event Horizon:} The appearance of
an event horizon becomes evident in the
2-dimensional model of the Alcubierre space-time, with $\rho=0$
\cite{Hiscock,Clark,Gonz}. The axis of symmetry coincides with the
line element of the spaceship.

The metric, eq.$(1)$, reduces to :
\begin{equation}
ds^2=-(1-v^2 f^2)dt^2-2vfdzdt+dz^2
\end{equation}

For simplicity, we consider the velocity of the bubble constant,
$v(t)=v_{b}$. With $\rho=0$, we have $r=[(z-v_{b}t)^2]^{1/2}$. If
$z>v_{b}t$, we consider the following transformation:
$r=(z-v_{b}t)$. Note that the metric components of eq.$(8)$ only
depend on $r$, which may be adopted as a coordinate.

Using the transformation, $dz=dr+v_{b}dt$, the metric, eq.$(8)$ is
given by:
\begin{equation}
ds^2=-A(r)\left[dt-\frac{v_{b}(1-f(r))}{A(r)}dr\right]^2+\frac{dr^2}{A(r)}
\end{equation}

The function $A(r)$, designated by the Hiscock function, is given
by:
\begin{equation}
A(r)=1-v_{b}^2(1-f(r))^2
\end{equation}

It's possible to represent the metric, eq.$(9)$, in a diagonal
form, using a new time coordinate:
\begin{equation}
d\tau=dt-\frac{v_{b}(1-f(r))}{A(r)}dr
\end{equation}
with which eq.$(9)$ reduces to:
\begin{equation}
ds^2=-A(r)d\tau^2+\frac{dr^2}{A(r)}
\end{equation}

This form of the metric is manifestly static. The $\tau$
coordinate has an immediate interpretation in terms of an observer
on board of a spaceship: $\tau$ is the proper time of the
observer, because $A(r)\rightarrow 1$ in the limit $r\rightarrow
0$.

We verify that the coordinate system is valid for any value of
$r$, if $v_{b}<1$. If $v_{b}>1$, we have a coordinate singularity
and an event horizon at the point $r_{0}$ in which
$f(r_{0})=1-\frac{1}{v_{b}}$ and $A(r_{0})=0$.

\section{The 2-dimensional Krasnikov solution}

The Krasnikov metric has the interesting property that although
the time for a one-way trip to a distant destination cannot be
shortened, the time for a round trip, as measured by clocks at the
starting point (e.g. Earth), can be made arbitrarily short, as
will be demonstrated below.

The 2-dimensional metric is given by:
\begin{equation}
ds^2=-(dt-dx)(dt+k(t,x)dx)
\end{equation}
where:
\begin{equation}
k(t,x)=1-(2-\delta)
\theta_{\varepsilon}(t-x)\left[\theta_{\varepsilon}(x)-
\theta_{\varepsilon}(x+\varepsilon-D)\right]
\end{equation}
in which $\delta$ and $\varepsilon$ are arbitrarily small positive
parameters. $\theta_{\varepsilon}$ denotes a smooth monotone
function:
\[
\theta_{\varepsilon}(\xi)=\left\{ \begin{array}{ll}
                     1,   & {\rm if}\; \xi>\varepsilon, \\
                     0,    & {\rm if}\; \xi<0.
                     \end{array}
                     \right.
\]

There are three distinct regions in the Krasnikov two-dimensional
spacetime, which we shall summarize in the following manner.

{\bf The outer region:} The outer region is given by the following
set:
\begin{equation}
\{x<0\}\cup\{x>D\}\cup\{x>t\}
\end{equation}

The metric is flat, $k=1$, and reduces to the Minkowski spacetime.
Future light cones are generated by the vectors:
\[
\left\{ \begin{array}{ll}
                     r_{O}=\partial_{t}+\partial_{x} \\
                     l_{O}=\partial_{t}-\partial_{x}.
                     \end{array}
                     \right.
\]

{\bf The inner region:} The inner region is given by the following
set:
\begin{equation}
\{x<t-\varepsilon\}\cap\{\varepsilon<x<D-\varepsilon\}
\end{equation}

This region is also flat, $k=\delta-1$, but the light cones are
{\it more open}, being generated by the following vectors:
\[
\left\{ \begin{array}{ll}
                     r_{I}=\partial_{t}+\partial_{x} \\
                     l_{I}=-(1-\delta)\partial_{t}-\partial_{x}.
                     \end{array}
                     \right.
\]

{\bf The transition region:} The transition region is a narrow
curved strip in spacetime, with width $\sim \varepsilon$. Two
spatial boundaries exist between the inner and outer regions. The
first lies between $x=0$ and $x=\varepsilon$, for $t>0$. The
second lies between $x=D-\varepsilon$ and $x=D$, for $t>D$. It is
possible to view this metric as being produced by the crew of a
spaceship, departing from point $A$ ($x=0$), at $t=0$, travelling
along the $x$-axis to point $B$ ($x=D$) at a speed, for
simplicity, infinitesimally close to the speed of light, therefore
arriving at $B$ with $t\approx D$.

The metric is modified by changing $k$ from $1$ to $\delta-1$
along the $x$-axis, in between $x=0$ and $x=D$, leaving a
transition region of width $\sim \varepsilon$ at each end for
continuity. But, as the boundary of the forward light cone of the
spaceship at $t=0$ is $|x|=t$, it is not possible for the crew to
modify the metric at an arbitrary point $x$ before $t=x$. This
fact accounts for the factor $\theta_{\varepsilon}(t-x)$ in the
metric, ensuring a transition region in time between the inner and
outer region, with a duration of $\sim \varepsilon$, lying along
the wordline of the spaceship, $x\approx t$.

\subsection{Superluminal travel within the Krasnikov tube}

The properties of the modified metric with $\delta-1\leq k \leq 1$
can be easily seen from the factored form of $ds^2=0$. The two
branches of the forward light cone in the $(t,x)$ plane are given
by $\frac{dx}{dt}=1$ and $\frac{dx}{dt}=-k$.

The inner region, with $k=\delta -1$, is flat because the metric,
eq.$(13)$, may be cast into the Minkowski form, applying the
following coordinate transformations:
\begin{equation}
dt'=dt+\left( \frac{\delta}{2}-1 \right) dx
\end{equation}
\begin{equation}
dx'=\left( \frac{\delta}{2} \right) dx
\end{equation}
The transformation is singular at $\delta=0$, i.e., $k=-1$. Note
that the left branch of the region is given by
$\frac{dx'}{dt'}=-1$.

From the above equations, one may easily deduce the following
expression:
\begin{equation}
\frac{dt}{dt'}=1+\left( \frac{2-\delta}{\delta}
\right)\frac{dx'}{dt'}
\end{equation}
For an observer moving along the positive $x'$ and $x$ directions,
with $\frac{dx'}{dt'}<1$, we have $dt'>0$ and consequently $dt>0$,
if $0<\delta \leq 2$. However, if the observer is moving
sufficiently close to the left branch of the light cone, given by
$\frac{dx'}{dt'}=-1$, eq.$(19)$ provides us with
$\frac{dt}{dt'}<0$, for $\delta< 1$. Therefore $dt<0$, the
observer traverses backward in time, as measured by observers in
the outer region, with $k=1$.

The superluminal travel analysis is as follows. Imagine a
spaceship leaving star $A$ and arriving at star $B$, at the
instant $t\approx D$. The crew of the spaceship modify the metric,
so that $k\approx -1$, for simplicity, along the trajectory.

Now suppose the spaceship returns to star $A$, travelling with a
velocity arbitrarily close to the speed of light, i.e.,
$\frac{dx'}{dt'}\approx -1$. Therefore, from eqs$(17)$-$(18)$, one
obtains the following relation:
\begin{equation}
v_{return}=\frac{dx}{dt}\approx
-\frac{1}{k}=\frac{1}{1-\delta}\approx 1
\end{equation}
and $dt<0$, for $dx<0$.

The return trip from star $B$ to $A$ is done in an interval of
$\Delta t_{return}=-D/v_{return}=D/(\delta -1)$. The total
interval of time, measured at $A$, is given by $T_A =D+\Delta
t_{return}=D \delta$. For simplicity, consider $\varepsilon$
negligible.

Superluminal travel is implicit, because $|\Delta t_{return}|<D$,
if $\delta >0$, i.e., we have a spatial spacetime interval between
$A$ and $B$. Note that $T_A$ is always positive, but may attain a
value arbitrarily close to zero, for an appropriate choice of
$\delta$.

\subsection{The 4-dimensional generalization}

The metric in the 4-dimensional spacetime, written in cylindrical
coordinates, is given by \cite{Everett}:
\begin{equation}
ds^2=-dt^2+(1-k(t,x,\rho))dx dt+k(t,x,\rho)dx^2+d\rho^2+\rho^2
d\phi^2
\end{equation}
with:
\begin{equation}
k(t,x,\rho)=1-(2-\delta)\theta_{\varepsilon}(\rho_{max}-\rho)
\theta_{\varepsilon}(t-x-\rho)[\theta_{\varepsilon}(x)-
\theta_{\varepsilon}(x+\varepsilon-D)]
\end{equation}

For $t\gg D+\rho_{max}$ one has a tube of radius $\rho_{max}$
centered on the $x$-axis, within which the metric has been
modified. This structure is designated by the {\it Krasnikov
tube}. In contrast with the Alcubierre spacetime metric, the
metric of the Krasnikov tube is static once it has been created.

The stress-energy tensor element $T_{tt}$ given by:
\begin{equation}
T_{tt}=\frac{1}{32
\pi(1+k)^2}\left[-\frac{4(1+k)}{\rho}\frac{\partial k}{\partial
\rho}+3\left(\frac{\partial k}{\partial
\rho}\right)^2-4(1+k)\frac{\partial^2 k}{\partial \rho^2}\right]
\end{equation}
can be shown to be the energy density measured by a static
observer, and violates the WEC in a certain range of $\rho$, i.e,
$T_{\mu\nu}U^{\mu}U^{\nu}<0$.

To verify the violation of the WEC, let us evaluate the energy
density in the middle of the tube and at a time long after it's
formation, i.e., $x=D/2$ and $t\gg x+\rho +\varepsilon $,
respectively. In this region we have $\theta_{\varepsilon}(x)=1$,
$\theta_{\varepsilon}(x+\varepsilon-D)=0$ and
$\theta_{\varepsilon}(t-x-\rho)=1$. With this simplification the
form function, eq.$(22)$, reduces to:
\begin{equation}
k(t,x,\rho)=1-(2-\delta)\theta_{\varepsilon}(\rho_{max}-\rho)
\end{equation}

Consider the following specific form for
$\theta_{\varepsilon}(\xi)$ \cite{Everett}:
\begin{equation}
\theta_{\varepsilon}(\xi)=\frac{1}{2}\left \{ \tanh \left
[2\left(\frac{2\xi}{\varepsilon}-1 \right )\right]+1 \right \}
\end{equation}
so that the above form function is given by:
\begin{equation}
k=1-\left (1-\frac{\delta}{2}\right)\left \{ \tanh \left
[2\left(\frac{2\xi}{\varepsilon}-1 \right )\right]+1 \right \}
\end{equation}

Choosing the following values for the parameters: $\delta =0.1$,
$\varepsilon =1$ and $\rho_{max}=100\varepsilon =100$, the
negative character of the energy density is manifest in the
immediate inner vicinity of the tube wall, as shown in fig.$(3)$.
\begin{figure}
\centering
\includegraphics[width=7cm]{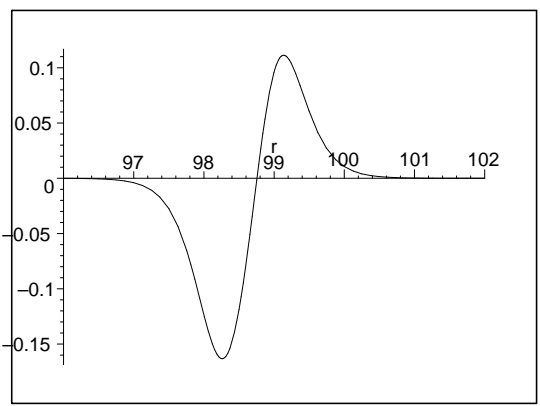}
\caption{Graph of the energy density, $T_{tt}$, as a function of
$\rho$ at the middle of the Krasnikov tube, $x=D/2$, and long
after it's formation, $t\gg x+\rho +\varepsilon $. We consider the
following values for the parameters: $\delta =0.1$, $\varepsilon
=1$ and $\rho_{max}=100\varepsilon =100$. \label{fig3}}
\end{figure}

\section{Superluminal travel requires the violation of the WEC}

It is simpler to apply global techniques and the topology of space
for a definition of superluminal travel. The following treatment
is based on work by Ken Olum \cite{Olum}.

A path, $P$, is defined along which a propagating signal travels
further than a signal on any nearby path, in the same interval of
externally defined time. Spacelike two-surfaces are constructed
around the origin and destination points of the path, $P$. The
spacetime metric is arranged so that a causal path exists between
the origin and destination points, $A$ and $B$, respectively, but
there are no other causal paths that connect the two-surfaces.
Both two-surfaces, $\Sigma_A$ and $\Sigma_B$, are composed of a
one-parameter family of spacelike geodesics through the respective
origin and destination points.

Formally, a causal path, $P$, is superluminal from an origin
point, $A$, to a destination point, $B$, only if it satisfies the
following condition.

\subsubsection{Superluminal Condition:}

There exists 2-surfaces $\Sigma_{A}$ around $A$ and $\Sigma_{B}$
around $B$ such that:

$(i)$ if $p\in \Sigma_{A}$ then a spacelike geodesic lying in
$\Sigma_{A}$ connects $A$ to $p$, and similarly for $\Sigma_{B}$,
and,

$(ii)$ if $p\in \Sigma_{A}$ and $q\in \Sigma_{B}$, then $q\in
J^{+} (p)$, i.e., $q$ is in the causal future of $p$, if and only
if $p=A$ and $q=B$.

\subsubsection{General considerations:}

Let there be a path $P$ satisfying the above condition, and
suppose that the generic condition holds on $P$ (recall that the
generic condition states that the path $P$ contains a point in
which $k_{[a}R_{b]cd[e}k_{f]}k^{c}k^{d}\neq0$ is satisfied, where
$k$ is the tangent vector to the geodesic). With these
assumptions, it can be shown that the WEC must be violated,
somewhere along $P$.

Note that $P$ must be a null geodesic.

{\it Proof}: If $P$ is not a geodesic it can be varied to make a
timelike path  from $A$ to $B$. Let $O$ be an open neighborhood of
$B$ contained in $\Sigma_{B}$. If $P$ is timelike anywhere, then
it can be varied to make a timelike path from $A$ to points of $O$
other than $B$, contradicting the Superluminal Condition.

Let $k$ be the tangent vector to the geodesic $P$. $k$ must be
normal to the surface $\Sigma_{A}$, otherwise there would be
points on $\Sigma_{A}$ in the past of points on $P$. Similarly for
$\Sigma_{B}$.

We define a congruence of null geodesics with an affine parameter
$u$, normal to $\Sigma_{A}$, and extend $k$ to be the tangent
vector at each point of the congruence.

There is no point $x\in P$ that is conjugate to $\Sigma_{A}$.

{\it Proof}: If $x$ were an interior point of $P$ then it would be
possible to deform $P$ into a timelike path \cite{Wald}. If $x=B$,
then different geodesics of the congruence would all end at $B$ or
points in an open neighborhood close to $B$ contained in
$\Sigma_{B}$. These geodesics would have different tangent vectors
to $\Sigma_{B}$. Therefore no point on $P$ is conjugate to
$\Sigma_{A}$.

The expansion of the geodesic congruence is given by
$\hat{\theta}=k^{m}_{\;\;\;;m}$\,, where $m$ runs over two
orthogonal directions normal to $k$. At $A$ we use directions that
lie in $\Sigma_{A}$ and at $B$ we use directions that lie in
$\Sigma_{B}$. Since $\Sigma_{A}$ is extrinsically flat at $A$, the
geodesics are initially parallel, therefore $\hat{\theta}=0$.

The evolution of the expansion, $\hat{\theta}$, is given by the
Raychaudhuri equation for null geodesics:
\begin{equation}
\frac{d\hat{\theta}}{du}=-R_{\mu\nu}k^{\mu}k^{\nu}+
2\hat{\omega}^2-2\hat{\sigma}^2-\frac{1}{2}\hat{\theta}^2
\end{equation}
in which $\hat{\omega}$ is the twist, $\hat{\sigma}$ the shear and
$R_{\mu\nu}$ is the Ricci tensor.

Since there are no conjugate points, $\hat{\theta}$ is
well-defined along $P$. We also have $\hat{\omega}_{\mu\nu}=0$,
because the congruence is (locally) hypersurface orthogonal,
according to the dual formulation of Frobenius' theorem
\cite{Wald}.

If the WEC holds, then by continuity the null energy condition
(NEC) will also be satisfied. The NEC is given by
$T_{\mu\nu}k^{\mu}k^{\nu}\geq 0$, for all null $k^\mu$. Using
Einstein's equation, we obtain $R_{\mu\nu}k^{\mu}k^{\nu}=8 \pi
T_{\mu\nu}k^{\mu}k^{\nu}$. Thus if the WEC is satisfied, then
$R_{\mu\nu}k^{\mu}k^{\nu}\geq 0$ and therefore
$\frac{d\hat{\theta}}{du}\leq 0$. From the generic condition,
$k_{[a}R_{b]cd[e}k_{f]}k^{c}k^{d}\neq0$ on a point along $P$,
$\hat{\sigma}$ cannot vanish everywhere. Recall that
$\hat{\theta}=0$ at $A$. Thus, the WEC implies, that at $B$, we
have:
\begin{equation}
\hat{\theta}<0.
\end{equation}

\subsubsection{Weak energy condition violation:}

If we can prove that the expansion obeys the inequality,
$\hat{\theta}\geq 0$, at $B$, then the WEC is violated somewhere
along $P$.

Firstly, it's important to establish a basis for vectors at $B$.
Let $E_1$ and $E_2$ be orthonormal vectors tangent to $\Sigma_B$
at $B$, and let $E_3$ be a unit spacelike vector orthonormal to
$E_1$ and $E_2$, with $g(k, E_3)>0$. Let $E_4$ be the unit
future-directed timelike vector orthogonal to $E_1$, $E_2$ and
$E_3$. Using this basis, normal Riemannian coordinates are
established near $B$, so that the $2$-surface $\Sigma_B$ consists
of points with $t=z=0$.

Let $\lambda(s)$ be a smooth curve on $\Sigma_A$, with
$\lambda(0)=A$. Let $\lambda(s,u)$ be the point an affine distance
$u$ along the null geodesic from $\lambda(s)$. Each geodesic will
eventually pass near $B$ and will cross the hypersurface with
$t=0$. This crossing point is called $\lambda'(s)$, and the length
of the vectors $k$ on $\Sigma_A$ are adjusted, so that
$\lambda(s,1)=\lambda'(s)$.

The $z$ coordinate of $\lambda'(s)$ is negative, otherwise points
on $\Sigma_{B}$ $(z=t=0)$ would be the future of points of the
geodesics from $\Sigma_A$, contradicting the Superluminal
Condition.

Let $Z$ be the tangent vector to $\lambda(s,u)$ in the $s$
direction. By construction, $k^{\mu} Z_{\mu}=0$ on $\Sigma_A$,
which is constant along each geodesic \cite{Wald,Hawking}, so that
$k^{\mu} Z_{\mu}=0$ is verified everywhere.

Following along $\lambda'(s)$, from $B$, we have:
\begin{equation}
0=\frac{d}{ds}(k^{\mu} Z_{\mu})=(k^{\mu} Z_{\mu})_{;\nu}
Z^{\nu}=k^{\mu}_{\;\;;\nu} Z_{\mu} Z^{\nu}+k^{\mu} Z_{\mu;\nu}
Z^{\nu}
\end{equation}

The only non-vanishing components of $k$ are $k^3$ and $k^4$.
Since $\lambda'(s)$ lies in the $t=0$ hypersurface, we have
$Z^4=0$ everywhere, so that the only contribution to $k^{\mu}
Z_{\mu;\nu}$ at $B$ is from $\mu=3$. Therefore, from the above
relation, we have:
\begin{equation}
k^{\mu}_{\;\;;\nu} Z_{\mu} Z^{\nu}=-k^{3} Z_{3;\nu} Z^{\nu}
\end{equation}

But at $B$, we have $Z_3=0$. We see that $Z_{3;\nu} Z^{\nu}\leq
0$, otherwise the $z$ coordinate of $\lambda'(s)$ would become
positive. By construction, $k^3>0$, so that $k^{3} Z_{3;\nu}
Z^{\nu}\leq 0$ and $k^{\mu}_{\;\;;\nu} Z_{\mu} Z^{\nu}\geq 0$.

The congruence of geodesics provides a map from tangent vectors to
$\lambda(s)$ at $A$ to tangent vectors to $\lambda'(s)$ at $B$. As
there are no conjugate points, the map is non-singular and can be
inverted. Choices of $\lambda(s)$ can be found so that $Z=E_1$ or
$Z=E_2$, thus $k^1_{\;\;;1} \geq 0$ and $k^2_{\;\;;2} \geq 0$,
respectively, so that:
\begin{equation}
\hat{\theta}=k^{m}_{\;\;\;;m}\geq 0
\end{equation}
contradicting $\hat{\theta}< 0$.

\subsubsection{Olum's superluminal theorem:}

Any spacetime that admits superluminal travel on some path $P$
(according to Olum's definition of the Superluminal Condition) and
that satisfies the generic condition on $P$, must also violate the
WEC at some point of $P$.

\subsection{Applications to the Casimir effect}

It was already mentioned that although classical forms of matter
obey the energy conditions, these are violated by certain quantum
fields, amongst which we may refer to the quantized scalar and
fermionic fields, the Casimir and the Topological Casimir Effect,
squeezed vacuum states, the Hawking evaporation, the
Hartle-Hawking vacuum, cosmological inflation, etc.

It is interesting to apply the Superluminal Condition to the
Casimir effect \cite{Olum}. The quantum expectation value of the
electromagnetic stress-energy tensor between circular conducting
plates is:
\begin{equation}
T_{\mu\nu}=\frac{\pi^2}{720 d^4}diag(-1,1,1,-3)
\end{equation}

For a geodesic travelling in the $z$-direction, we have:
\begin{equation}
R_{\mu\nu}k^{\mu}k^{\nu}=-\frac{2 \pi^2}{45 d^4}<0
\end{equation}

Let $\Sigma_{A}$ be the lower plate and $\Sigma_{B}$ be the upper
plate. Assuming that all the geodesics are initially parallel, so
that $\hat{\theta}=0$ at $A$. We have $\hat{\sigma}=0$, by
symmetry, and $\hat{\omega}=0$, because the congruence is
hypersurface orthogonal. The Raychaudhuri equation reduces to:
\begin{equation}
\frac{d\hat{\theta}}{du}=-R_{\mu\nu}k^{\mu}k^{\nu}>0
\end{equation}

This inequality shows that the geodesics around $P$ are defocused.
Thus the geodesic $P$ travels further in the $z$-direction, by the
same same $t$, than neighbouring geodesics, in which the
Superluminal Condition is satisfied.

It is also important to note that the above analysis is probably
not complete, because the mass of the plates have not been taken
into account.

\section{Quantum Inequality and applications}

Intensive research has been going on into the violation of the
energy conditions. It is interesting to note the pioneering work
by Ford in the late 1970's on a new set of energy constraints
\cite{ford1}, which led to constraints on negative energy fluxes
in 1991 \cite{ford2}. These eventually culminated in the form of
the Quantum Inequality (QI) applied to energy densities, which was
introduced by Ford and Roman in 1995 \cite{F&R1}.

The QI was proven directly from Quantum Field Theory, in
four-dimensional Minkowski spacetime, for free quantized, massless
scalar fields, and takes the following form:
\begin{equation}
\frac{\tau_0}{\pi}\int_{-\infty}^{+\infty}\frac{\langle{T_{\mu\nu}U^{\mu}U^{\nu}}\rangle}
{\tau^2+{ \tau^2_0}}d\tau\geq-\frac{3}{32\pi^2\tau^4_0}, \label{1}
\end{equation}
in which, $U^\mu$ is the tangent to a geodesic observer's
wordline; $\tau$ is the observer's proper time and $\tau_0$ is a
sampling time. The expectation value $\langle\rangle$ is taken
with respect to an arbitrary state $|\Psi\rangle$. One does not
average over the entire wordline of the observer, as in the
averaged energy conditions, but weights the integral with a
sampling function of characteristic width, $\tau_0$. The
inequalities limit the magnitude of the negative energy violations
and the time for which they are allowed to exist. The basic
applications to curved spacetimes is that these appear flat if
restricted to a sufficiently small region.

Using the restrictions imposed by the QI to wormholes \cite{Roman}
and the warp drive \cite{PfenningF}, it was verified that the
throat size of the wormholes and the Alcubierre bubble wall are
extremely thin, i.e., only slightly larger than the Planck length.
It was also verified that the energy involved to support the
Alcubierre bubble and the Krasnikov tube are probably not
physically plausible, for they are extraordinary large. For
example, considering the mass of a typical galaxy, $M_{Milky
Way}\approx 10^{12} M_{Sun}=2\times 10^{42} kg$, the energy
necessary to support the Alcubierre bubble is $E\leq-5,5\times
10^{21} M_{Milky Way}\times v_b$, which is of the order $10^{10}$
times the total mass of the Universe. In the opposite regime, for
microscopic Alcubierre bubbles, of the order of the Compton length
of an electron, the negative energy is of the order $E\sim -10^{4}
M_{Sun}$. Due to these enormous amounts of exotic matter, van den
Broeck proposed a slight modification of the Alcubierre metric
which ameliorates considerably the conditions of the warp drive
\cite{Broeck1}.

Considering the applications of the QI to the above-mentioned
solutions, one may, rightly so, conclude that these solutions are
not physically plausible. However, there are a series of
considerations that can be applied to the QI \cite{Lobo}. Firstly,
the QI is only of interest if one is relying on quantum field
theory to provide the exotic matter to support the solutions
above-mentioned. But there are classical systems (non-minimally
coupled scalar fields) that violate the null and the weak energy
conditions \cite{B&V}, whilst presenting plausible results when
applying the QI. Secondly, even if one relies on quantum field
theory to provide exotic matter, the QI does not rule out the
existence of the considered solutions, although they do place
serious constraints on the geometry.

Despite of the impressive work done by Ford and Roman, namely the
deduction of the QI and all it´s applications
\cite{Everett,Roman,PfenningF,PFord} the current version of the QI
is certainly not the last word on the subject of exotic matter and
the energy condition violations.

\section{Conclusion}

It does seem to suggest that if one adopts a conservative view,
and impose the WEC, Olum's theorem prohibits superluminal travel.
As was mentioned in the introduction the present work serves as a
bridge to ongoing research on spacetimes which generate closed
timelike curves. An extension of Olum's theorem to these
spacetimes is the next step, or the generalization and
modification of his superluminal definition. This is not easily
accomplished because most of the definitions adopted in the causal
structure of spacetime \cite{Wald,Hawking} break down in the
presence of CTCs.

%INDEX%%%%%%%%%%%%%%%%%%%%%%%%%%%%%%%%%%%%%%%%%%%%%%%%%%%%%%%%%%%%%%%
% Please check with the editor of your book whether he plans to
% include a "mutual" subject index - if so, please code your entries
% in the standard syntax. For your own purposes you may print your
% "personal" index by using the following commands:
%
%\clearpage
%\addcontentsline{toc}{section}{Index}
%\flushbottom
%\printindex
%%%%%%%%%%%%%%%%%%%%%%%%%%%%%%%%%%%%%%%%%%%%%%%%%%%%%%%%%%%%%%%%%%%%%

\end{document}